# Field-effect at electrical contacts to two-dimensional materials

Yan Sun[1], Alvin Tang[2], Ching-Hua Wang[2], Yanqing Zhao[1], Mengmeng Bai[1], Shuting Xu[1], Zheqi Xu[1], Tao Tang[3], Sheng Wang[4], Chenguang Qiu[4], Kang Xu[5], Xubiao Peng[1], Junfeng Han[1], Eric Pop[2], and Yang Chai[5], Yao Guo[1*]

[1]School of Physics, Beijing Institute of Technology, Beijing 100081, China.

[2]Department of Electrical Engineering and Stanford SystemX Alliance, Stanford University,

Stanford, California 94305, United States.

[3]Advanced Manufacturing EDA Co., Ltd, Shanghai, 201204, China.

[4]Key Laboratory for the Physics and Chemistry of Nanodevices, Department of Electronics,

Peking University, Beijing 100871, China.

[5]Department of Applied Physics, The Hong Kong Polytechnic University, Hong Kong, China.

* Corresponding author: yaoguo@bit.edu.cn

**ABSTRACT**

The inferior electrical contact to two-dimensional (2D) materials is a critical challenge for their application in post-silicon very large-scale integrated circuits. Electrical contacts were generally related to their resistive effect, quantified as contact resistance. With a systematic investigation, this work demonstrates a capacitive metal-insulator-semiconductor (MIS) field-effect at the electrical contacts to 2D materials: the field-effect depletes or accumulates charge carriers, redistributes the voltage potential, and give rise to abnormal current saturation and nonlinearity. On the one hand, the current saturation hinders the devices' driving ability, which can be eliminated with carefully engineered contact configurations. On the other hand, by introducing the nonlinearity to monolithic analog artificial neural network circuits, the circuits' perception ability can be significantly enhanced, as evidenced using a COVID-19 critical illness prediction model. This work provides a comprehension of the field-effect at the electrical contacts to 2D materials, which is fundamental to the design, simulation, and fabrication of electronics based on 2D material.

**INTRODUCTION**

With their abundance and the rich variety of electronic properties, the family of atomically thin two-dimensional (2D) materials is a promising choice in the next generation very-large-scale-integration (VLSI) technology.[1-7] Electronics devices based on atomically thin 2D materials such as transition metal sulfides, black phosphorus, graphene, *etc.,* have been intensively studied during the past decade. 2D material-based electronics show great potential for building both traditional von Neumann computing architecture and state-of-art non-von Neumann in-memory-computing architecture.[8-17] However, the inferior electrical contacts greatly hinder the performance of 2D material-based electronic devices.[18-22] Multiple elaborate methods have been demonstrated to improve the electrical contact to 2D materials, such as low work function metallization,[23] interface engineering,[24, 25] doping,[26-29] phase change transition,[30] and the recent van der Waals contacts.[31-33] Still, the electrical contact to 2D materials is yet unable to meet the current VLSI technology standard.[3] Herein continuous improvement of the electrical to 2D materials is required, and full fundamental comprehension of the electrical contacts to 2D materials is urgently needed.

In previous studies, researchers have generally focused on the resistive effect of electrical contacts, quantified as the contact resistance.[16, 34] In this work, we investigate the electrical contacts to 2D materials from a different physical perspective and demonstrate that for the atomically thin 2D materials, the electrical contacts contribute not only the resistive effect but also a significant capacitive metal-insulator-semiconductor (MIS) field-effect. The capacitive MIS field-effect can deplete or accumulate the charge carriers, redistribute the voltage potential, and give rise to unusual current saturation and nonlinearity. Based on a thorough understanding, one can purposively decide whether to eliminate such field-effect or use it. On the one hand, engineering on the electrical contact configuration can eliminate the field-effect and significantly improve the driving current of the 2D material based transistors. On the other hand, the nonlinearity introduced by the electrical contact to 2D materials can efficiently improve the perceptron ability of the non-von Neumann in-memory-computing circuit, as demonstrated by a COVID-19 critical illness prediction model.

## I. Principle of the Capacitive MIS Field-Effect at Electrical Contact

For concision, we first expound on the general principle of the capacitive MIS field-effect. Fig. 1(a) shows a simplified 2D material channel (monolayer $MoS_2$, for example) with two contacting metallic electrodes. As shown in the inset, the sidewall of the metallic electrode, the insulating dielectric (or vacuum), and the semiconducting channel form a fringe metal-insulator-semiconductor (MIS) structure. At the ends of the channel, the distance between the metallic sidewall and the channel is small (Ångstroms to nanometers). The capacitive coupling between the metallic sidewall and the channel can be very effective. Therefore the voltage drop between the metallic sidewall and the channel (which is $V_{M-S} = I_{DS} \cdot R_C$, $I_{DS}$ is the current, and $R_C$ is the contact resistance) can create a significant MIS field-effect. As shown in Figs. 1(a), we simulated the simplified device (monolayer $MoS_2$, $n$-type, donor density $6 \times 10^{12}$ cm$^{-2}$, intrinsic mobility 400 cm$^2$V$^{-1}$s$^{-1}$, channel length 500 nm, contact resistance $2R_C$ 1.3kΩ ) with technology computer-aided design (TCAD). Fig. 1(b) and Figs. 1(b) show the simulated electron density distribution in the channel. For $V_{DS} = 0$ V, the electron density is inconsistent with the donor doping density. With applied $V_{DS} = 1$ V, carriers deplete/accumulate at the Source/Drain end of the channel. For increased $V_{DS}$ up to 5 V, the depletion/accumulation is further strengthened. The depletion/accumulation is due to the field-effect coupled through the MIS structure, where the voltage drop $\pm V_{M-S}$ are applied between the Source/Drain electrode sidewall and the channel, respectively. The fully depleted region is non-conducting, correspondingly, a considerable part of voltage potential drops on the depletion region, as shown in Fig. 1(c). Considering this capacitive MIS field-effect, we plot the equivalent circuit in Fig. 1(d). Besides the Source/ Drain contact resistors ($R_C$) and the channel resistor ($R_{Ch}$), the circuit consists of two extra parasitic MIS field-effect transistors. The gates of the transistors correspond to the sidewalls of the electrodes. The $I_{DS}$-$V_{DS}$ curves of the equivalent circuit with various contact/channel resistance ($2R_C = R_{Ch}$) are shown in Fig. 1(e), which are in good accordance with the TCAD simulated $I_{DS}$-$V_{DS}$ curves of the constructed device with various doping density, as shown in Fig. 1(f). The current increases linearly with small $V_{DS}$ and saturates with large $V_{DS}$. The current saturation results from a negative feedback mechanism: with increased $V_{DS}$, the field-effect at the Source end can fully deplete the carriers through the atomically thin 2D material, dramatically decreases the conductivity, prevents the $I_{DS}$ from increasing, and in turn causes apparent current saturation.

We further carry comparative simulations by replacing the Source/Drain with the virtual zero-thickness electrodes, as shown in Figs. 1(c). In this case, the fringe capacitive MIS field-effect coupling is absent. Consequently, we do not see the carrier depletion/accumulation at the ends of the channel, as shown in Fig. 1(g) and Figs. 1(d). The voltage drop redistributes equably, and the $I$-$V$ curves maintain good linearity, as shown in Fig. 1(h) and 1(i). The comparative simulations confirm the field-effect originated from the MIS structure. The zero-thickness virtual electrodes neglect the apparent field-effect at the electrical contact to 2D materials, which should be used with extra caution in TCAD simulations.

The above analysis has proposed a field-effect at the electrical contacts to 2D materials, which can significantly deplete/accumulate carriers, change the voltage potential distribution, and give rise to apparent current saturation. For clarification, the above simulations have used the simplified device geometry with only one channel and two electrodes. However, the principle is universal and applies to practical 2D material devices with more complex geometries, as will be shown below. A similar analysis also applies to Schottky contacts,[35] where the field-effect widens the Schottky barrier width at the contact area, blocks field/thermal-field emission current, and causes a similar current saturation, as we specify detailedly in supplementary material II.

## II. Capacitive MIS field-effect at the electrical contacts to monolayer $MoS_2$

So far, we have theoretically proposed the significant capacitive MIS field-effect of the electrical contacts to ultra-thin 2D material channels. The question is, in practical cases, does this field-effect have a real impact on 2D material electronic devices? To address this question, we experimentally fabricated the back gate monolayer $MoS_2$ FETs. Here we use Ti as the contacting metal. The transfer curves and output *I-V* curves are shown in Fig. 2(a) and 2(b), respectively. We note that the *I-V* curves are linear at small $V_{DS}$ and the current saturates at large $V_{DS}$, which is in accordance with the simulation results in Fig. 1(e) and 1(f). Similar current saturation is observed not only with Ti/$MoS_2$ contacts,[19, 36] but also with Ni/$MoS_2$ contacts (see Figs. 3),[37] Au/$MoS_2$ contacts,[11] Ti/$WSe_2$ contacts,[38] Ag/$WSe_2$ contacts,[38] van der Waals $VSe_2$/$WSe_2$ contacts,[21] *etc*. In all the above cases, we find $V_G$-$V_T$>>$V_{DS}$ ($V_G$ is the gate voltage and $V_T$ is the threshold voltage), therefore the current saturation should not result from the common pinch-off of the channel. Previous studies have attributed such abnormal current saturation to several possible mechanisms such as carrier velocity saturation,[37, 39] self-heating effect,[36] or Schottky barriers,[19] *etc.*, while the field-effect at the electrical contact has not been recognized.

To address the origin of the abnormal current saturation, we designed and fabricated asymmetric FETs with an arc channel, as shown in Fig. 2(c). The inner radius $r_1$ and the outer radius $r_2$ are 0.33 μm and 1.0 μm, respectively. Therefore the contact resistance is asymmetric, with $R_{C1}$ prominently larger than $R_{C2}$ ($R_{C1}$: $R_{C2}$ ≈ $r_2$ / $r_1$ = 3, considering the relatively small transfer length)[20]. The transfer curve of the device is shown in Fig. 2(d). Two sets of *I-V* curves were obtained comparatively: in Fig. 2(e), the outer electrode was used as the Source, the inner electrode as Drain, and the IV curves are linear; while in Fig. 2(f), the inner electrode was used as the Source, the outer electrode as Drain, the current starts to saturate with $V_{DS}$ ~ 2 V. This is direct evidence that the current saturation is closely related to the electrical contacts, rather than the channel. When the inner electrode is the Source, the voltage drop at $I_{DS} \cdot R_{C1}$ is relatively large, the field-effect depletes the electrons effectively and causes current saturation. When the outer electrode is the Source, the voltage drop at $I_{DS} \cdot R_{C2}$ is relatively small, which is not large enough to fully depletes the electrons or causes current saturation. Should the current saturation be attributed to the carrier velocity saturation or self-heating effect, the current saturation behavior in Fig. 2(e) and Fig. 2(f) would be identical. Should the current saturation be attributed to the switching of the Schottky barrier,[19] the current saturation should happen when the inner/outer electrode is used as the Drain/Source respectively, which is contrary to the above observations. The mere Schottky barrier (without considering field-effect) dominated I-V curves should present increased differential conductance at large $V_{DS}$, due to the lowered Schottky barrier height and enhanced field/thermal-field emission with increased $V_{DS}$,[40-42] as shown in Figs. 2(a).

Based on the above analysis, it is shown that the capacitive MIS field-effect of the electrical contacts can dominate the electrical characteristics of practical monolayer $MoS_2$ devices, giving rise to current saturation at large $V_{DS}$. Note that the contacting angle between the metal and the channel might be slanted, as shown in supplementary material IV. The field-effect coupling with the slanted angle can be more robust and give rise to more apparent carrier depletion/accumulation, voltage potential redistribution, and current saturation, as shown in supplementary material V. We also note that the current saturation is closely related to the degeneration of the device, as shown in supplementary material VI, probably because the surface of Ti is oxidized into $TiO_x$ with a high dielectric constant (~18) and further strengthen the capacitive MIS field-effect.[43]

## III. Capacitive MIS field-effect at the electrical contacts to graphene

The concept of capacitive MIS field-effect for electrical contacts is general and should apply to a wide range of ultra-thin channel devices. Here we show how the capacitive MIS field-effect of the electrical contacts affects the characteristics of graphene FETs. Unlike semiconducting 2D materials with sizeable bandgaps, graphene is semi-metallic with zero bandgap. Considering the capacitive MIS field-effect, for the *n*-type graphene channel, electrons deplete at the Source end and accumulate at the Drain end at a small voltage drop (Fig. 3(a) left). The Source end may further reverse at a large voltage drop, and holes accumulate due to the bandgap absence (Fig. 3(a) right). We fabricated the arc channel graphene FETs with $r_1$ of 0.5 μm and $r_2$ of 1.5 μm (Figs. 7(a)), and the Dirac point extracted from the transfer curve is 8 V (Figs. 7(b). We compared the *I-V* curves with the outer/inner electrode as the Source or Drain alternated. The current is larger with the outer/inner electrode as the Source/Drain electrode (Fig. 3(b), left), and the derived differential conductance continually increases with $V_{DS}$ (Fig. 3(c), left, $V_G - V_{Dirac} = 52$ V, where $V_{Dirac}$ is the Dirac point voltage). In contrast, the current with the inner/outer electrode as the Source/Drain electrode is smaller (Fig. 3(b), right, compared in Figs. 7(c)), and the derived differential conductance first decreases and then increases with $V_{DS}$ (Fig. 3(c), right). Furthermore, the inversion point of $V_{DS}$ is monotonically dependent on the gating, as shown in Figs.7(e) and 7(g). The above observation is in good agreement with the principle of the capacitive MIS field-effect for the electrical contacts: the inner electrode with larger contact resistance dominates the outer electrode, when the inner electrode is used as the Drain, the electrons accumulate and the conductance continually increases; while when the inner electrode is used as the Source, the electrons firstly deplete and then the holes accumulate with increased $V_{DS}$, consequently, the conductance firstly decreases and then increases. The identical analysis applies to the *p*-type channel graphene, and similar results are observed, as is shown in Fig. 3(d), (e), (f), and Figs. 7(d), (f), (h). Therefore, the capacitive MIS field-effect of the electrical contacts applies to a wide range of 2D materials, including both semiconducting 2D materials and semi-metal 2D materials.

## IV. Engineering the Electrical Contact Configuration

The comprehension of field-effect at the electrical contact to 2D materials leads to an interesting question of whether to eliminate it or use it, depending on different application scenarios. For 2D material-based transistors, the field-effect at electrical contact saturate the ON current and hinders the devices' driving ability. The field-effect is critically dependent on the configuration. An elaborately designed electrical contact configuration should be able to eliminate the capacitive MIS field-effect effectively. As shown in Fig. 4(a) and Figs. 8, extended electrodes are realized by the butyl lithium induced 2H/1 T' phase change transition, so that the sidewall of the metallic electrode is kept far away from the semiconducting channel. Such configuration eliminates the capacitive MIS field-effect, and subsequently, the current is not saturated, as is shown in Fig. 4(b) and 4(c). Therefore, the efforts to improve the electrical contact to 2D materials should concentrate on not only the interface between the metallic electrode and 2D materials (to lower the contact resistance), but also the contact configuration (to eliminate the field-effect).

Using TCAD simulations, we further studied electrical monolayer $MoS_2$ short-channel transistors (10 nm gate length, 20 nm Source/Drain distance) with various contact configurations. The contact resistance is uniformly set to 800 Ω·µm. Fig. 4(d) shows two comparative transistors, the first transistor (upper) is with the common contact configuration, the second transistor (lower) is with the optimized contact configuration, including the extended electrodes and the low dielectric spacer, to suppress the coupling through the capacitive MIS field-effect. The transfer curves of the two transistors are compared in Fig. 4(e). The red lines are from the transistor with the common contacts, and the black lines are from that with the extended electrode contacts. Compared to the common contact configuration, the extended electrode increases the ON current from 96 µA·µm$^{-1}$ to 271 µA·µm$^{-1}$, the ON/OFF ratio from $1.8×10^4$ to $5×10^5$, suppresses the subthreshold swing (SS) from 95 mV/dec to 77 mV/dec, and the drain-induced-barrier-lowering (DIBL) from 89 mV·V$^{-1}$ to 7 mV·V$^{-1}$. Therefore, the optimized contact configuration eliminates the capacitive MIS field-effect effectively and improves the device performance significantly. We further propose other engineered contact configurations, such as the contacts with the oblique contacting angle or the contacts using graphene as the extended electrode,[44] as is shown detailedly in Figs. 9. The device performance with various contact configurations are summarized in Fig. 4(f) and Table 1. All of the three optimized electrical contact configurations dramatically improved the performance of monolayer MoS$_2$ short-channel transistors.

## V. Enhancing Perception Ability of Artificial Neural Network (ANN) Circuits

Artificial neural network (ANN) implanted in in-memory-computing circuits has great advantages in parallel computing acceleration and energy efficiency.[45-48] For ANN, neurons with nonlinear activation functions are essential. In the mathematical theories of ANN, the universal approximation theorem has demonstrated that given appreciate squashing activation function (for example, S-shape functions), a three-layer ANN perceptron consisting of an input layer, one hidden layer with an arbitrary number of neurons, and an output layer (Fig. 5(a)) can approximate any Lebesgue integrable function.[49-51] The S-shape tanh and sigmoid activation functions, which feature the saturating nonlinearity, are among the most commonly used activation functions in ANN. Analog/digital conversions with additional activation processors can achieve such nonlinearities, but they consume extra chiplet/die area and energy, and increase the complexity of the design and fabrication process.[52, 53]

In the above discussion, this work had demonstrated that the field-effect of the electrical contact to 2D materials can introduce saturating nonlinearity in IV curves. One can easily obtain a tanh-like or sigmoid-like IV curve with the $MoS_2$ or other 2D material based devices, as shown in Figs. 10. A three-layer in-memory-computing ANN perceptron circuit diagram is shown in Fig. 5(b). Such hybrid circuits can be achieved with the compact 3D integration shown in Figs. 11, where $MoS_2$ channels are nonlinear neurons, and random access memories are synapses. We simulated the ANN perceptron to approximate a COVID-19 early-stage prediction dataset model, as shown in Fig. 5(c). The inputs are ten clinical indexes, and the output is the predicted probability that the patient develops to critical illness within 5 days. Such prediction is essential for the triage treatment of COVID-19 patients, as detailedly described in an earlier paper by Zhong et al.[54] Here we use 3000 derivate patients' clinical indexes (input) and Zhong's prediction (output) for the training of the perception. As is shown in Fig. 5(d), with the nonlinearity introduced, the ANN reaches a lower loss than the linear regression in 45000 epochs. The regressed critical illness probability of 100 patients shows significantly improved accuracy than the linear regression model, with the average error decreased from 4.71 % to 2.87 %, as shown in Fig. 5(e) and 5(f). Therefore, the electrical contact introduced nonlinearity can significantly improve the accuracy of the in-memory-computing integrated ANN perceptron.

**VI. CONCLUSION**

This work proposes and demonstrates a field-effect at the electrical contacts to 2D materials, which has long been neglected in previous studies. The field-effect can deplete and accumulate the carriers, change the carrier densities and types, and introduce current saturation and nonlinearity. Based on such comprehension, on the one hand, we call for attention to the optimization of the electrical contact configuration for 2D material transistors, which increases the current driving ability of devices. On the other hand, we demonstrate that the nonlinearity introduced by the field-effect at the electrical contact can enhance the perception ability of the hybrid ANN circuits. This work provides a full comprehension of the electrical contacts to 2D materials, which is fundamental to the simulation, design, and fabrication of 2D material-based electronic devices.

## VII. METHODS

Simulations: The simulations are performed using the TCAD tool Sentaurus developed by Synopsys, Mountain View, CA, the US.

Experiments: For the Ti-$MoS_2$ and the Ti-graphene contacted devices, monolayer $MoS_2$ and graphene flakes were exfoliated onto a heavily doped silicon wafer (<0.005 Ω·cm) with 300 nm of thermally grown $SiO_2$. Electrodes were patterned using e-beam lithography, 10 nm Ti/ 50 nm Au was deposited to form the contacts using e-beam evaporation. The contact resistance of the Ti-$MoS_2$ contacts fabricated in our lab is sample/gate voltage-dependent, ranging from 20 kΩ·μm to 400 kΩ·μm.[22] Butyl lithium was used for the 2H to 1T' phase transition of $MoS_2$. For the Ni-$MoS_2$ contacted devices, monolayer $MoS_2$ was synthesized using the chemical vapor deposition and transferred onto a heavily doped silicon wafer with 60 nm of thermally grown $SiO_2$. Electrodes were patterned using e-beam lithography and 40 nm Ni was deposited using e-beam evaporation. After lift-off, the devices were immediately transferred into a vacuum probe station with the Keithley 4200 semiconductor characterization system. The devices were kept in a low vacuum environment ($1\times10^4$ Pa) for degeneration study in Figs. 6.

**SUPPLEMENTARY MATERIAL**

See the supplementary material for details on simulations, device characterization, TEM images, device fabrication, different contact configurations and ANN circuits.

**ACKNOWLEDGMENTS**

We thank Prof. H.-S. Philip Wong from Stanford University for the support of this work. We thank Prof. Zhiyong Zhang and Prof. Qing Chen from Peking University for useful discussions. Yao Guo thank Dr. I-Ting Wang for the valuable suggestions. This work was supported by the Beijing Institute of Technology Research Fund Program for Young Scholars and the National Natural Science Foundation of China grant 11804024.

**AUTHOR CONTRIBUTION**

Y. G. composed the work, conducted the experiments, and analyzed the data. Y. S., M. B., S. X. and T. T. conducted the TCAD simulations. Y. S., A. T., and C. W. fabricated the devices and conducted the measurements. Y. Z. and X. K. conducted the ANN simulations. Z. X., S. W., C. Q., X. P., J. H., E. P., and Y. C. contributed valuable discussions. Y. G. and Y. S. wrote the manuscript. Y. G. supervised the project. All authors provided critical feedback and helped shape the research, analysis, and manuscript.

**DATA AVAILABILITY**

All data needed to evaluate the conclusions in the paper are present in the paper and/or the Supplementary Material. Additional data related to this paper may be requested from the corresponding authors.

**Figures and Tables**

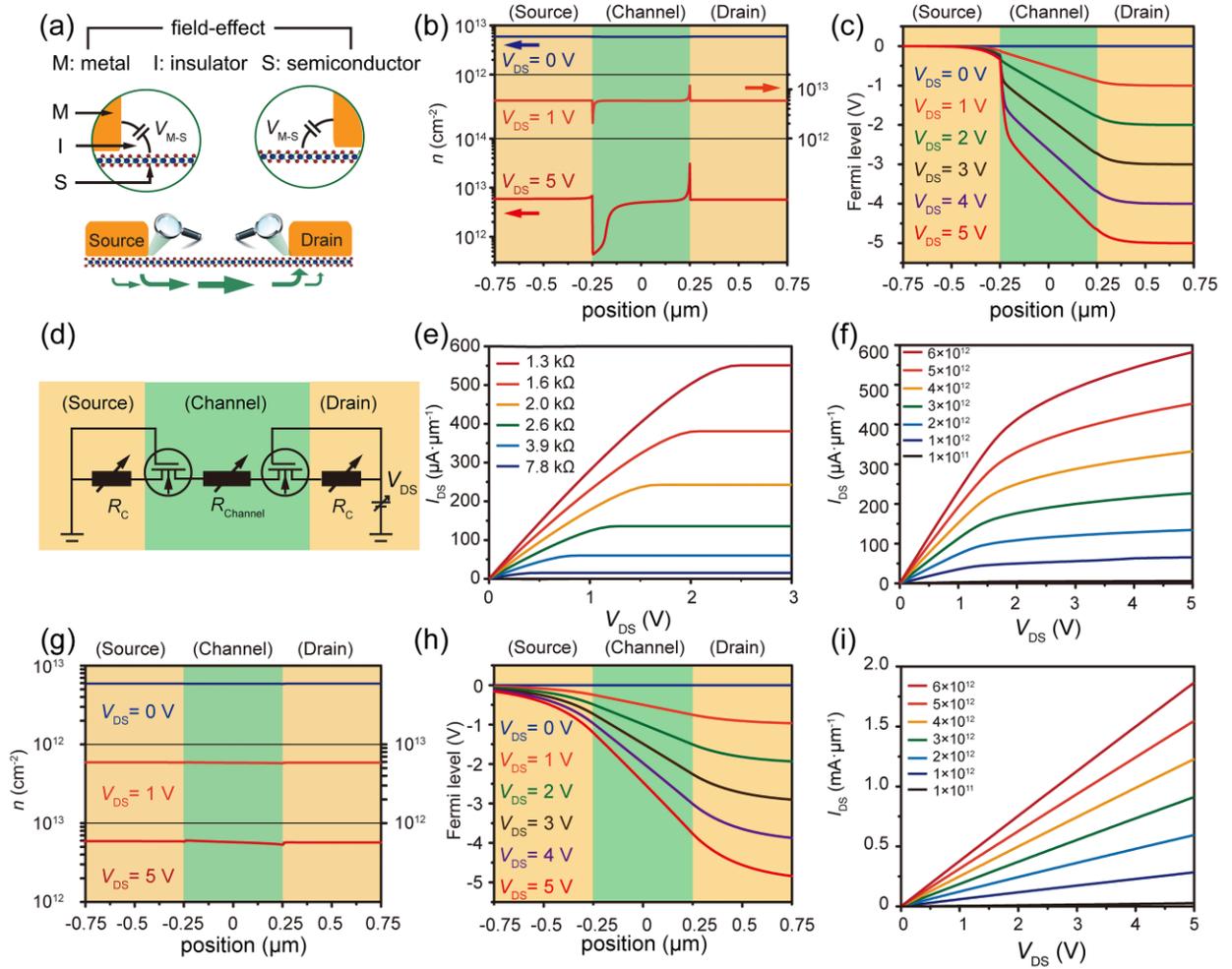

**FIG 1.** Principle of the capacitive MIS field-effect. (a) Capacitive MIS field-effect of electrical contacts coupled through the MIS structure. (b) Electron density and (c) voltage potential along the $MoS_2$ channel. (d) Equivalent circuit of devices considering the capacitive MIS field-effect. (e) $I_{DS}$-$V_{DS}$ curves of the equivalent circuit. The legends are the contact/channel resistance ($2R_C$=$R_{Ch}$). (f) $I_{DS}$-$V_{DS}$ curves of the TCAD simulated device, the legends are the doping density (cm$^{-2}$). (g) Electron density, (h) the voltage drop distribution, and (i) $I_{DS}$-$V_{DS}$ curves of the device without the capacitive MIS field-effect.

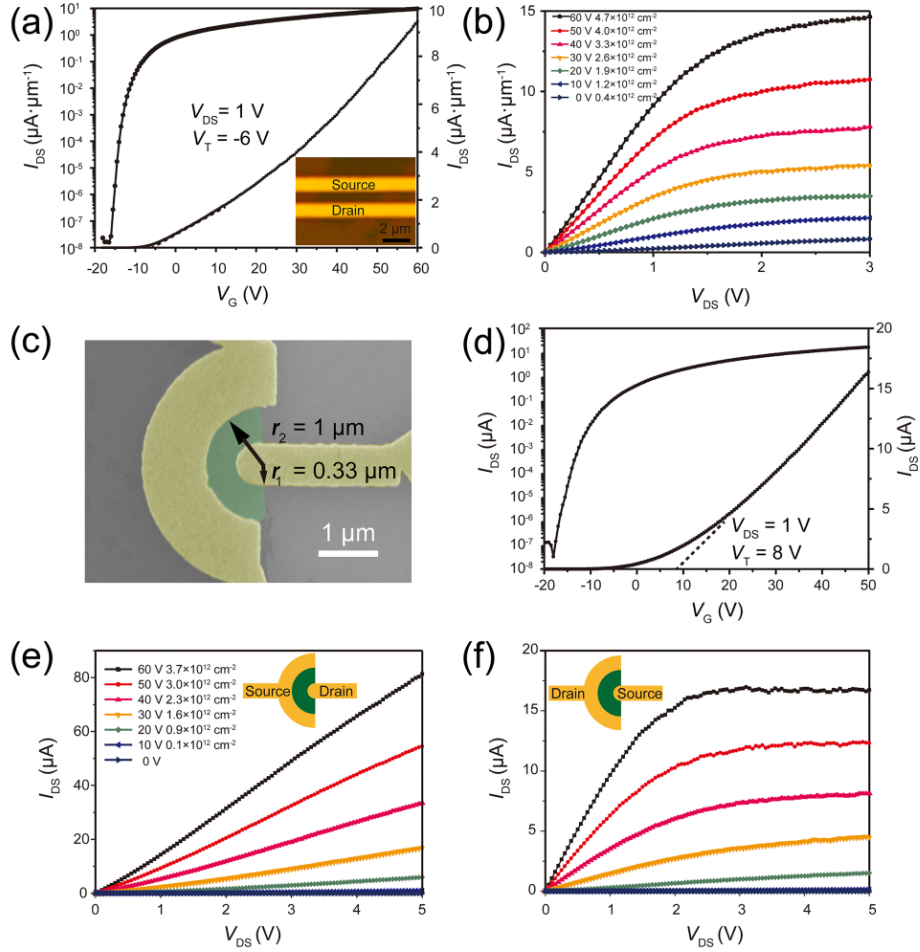

**FIG 2.** Capacitive MIS field-effect of electrical contacts to monolayer $MoS_2$. (a) Transfer curve and (b) *I-V* curves of the back gate monolayer $MoS_2$ FET. The inset in (a) is the microscope image of the device, the channel length is 0.8 μm. (c) SEM image of the asymmetric $MoS_2$ back gate FET with arc channel. (d) Transfer curve of the asymmetric $MoS_2$ back gate FET. (e) $I_{DS}$-$V_{DS}$ curves of the asymmetric $MoS_2$ back gate FET with the outer/inner electrode as the Source/Drain respectively. (f) $I_{DS}$-$V_{DS}$ curves of the asymmetric $MoS_2$ back gate FET with the inner/outer electrode as the Source/Drain respectively.

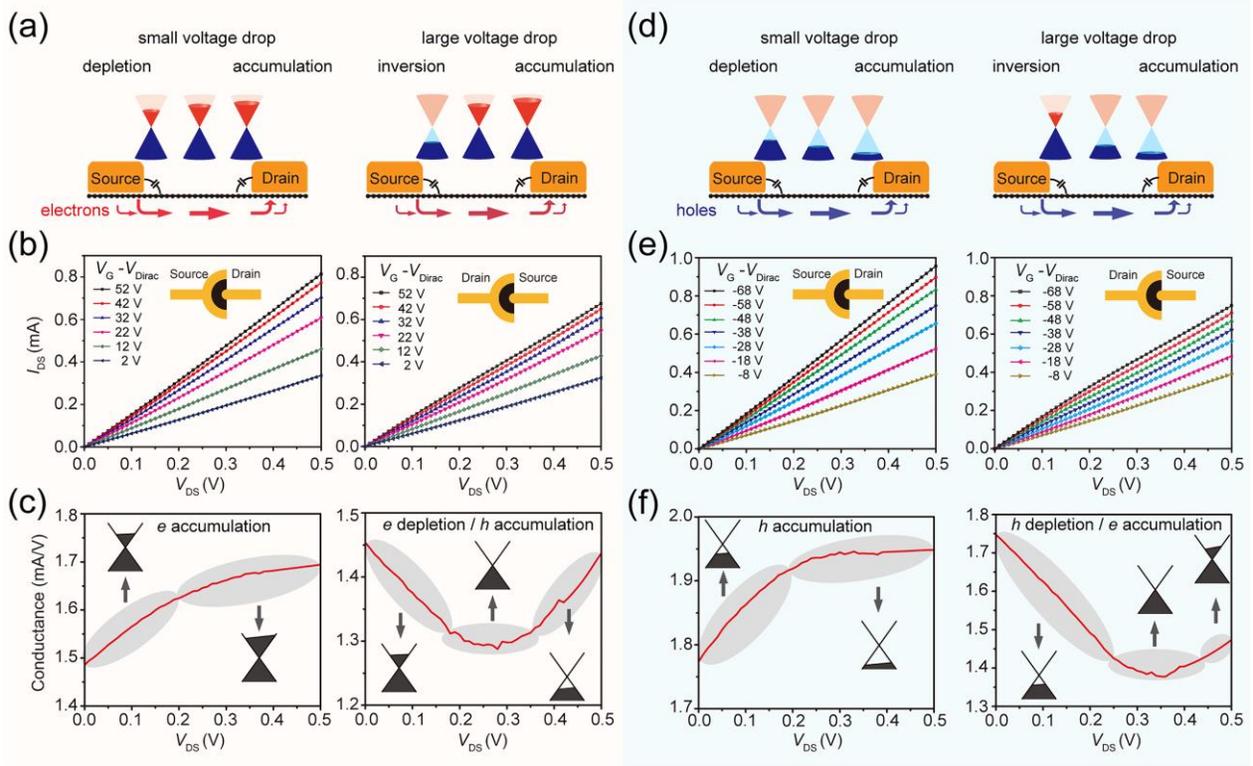

**FIG 3.** Capacitive MIS field-effect of the electrical contacts to graphene. Principle of the capacitive MIS field-effect for metal-graphene contacts, the depletion, accumulation, and inversion of the graphene channel: (a) *n*-region; (d) *p*-region. $I_{DS}$-$V_{DS}$ curves of the graphene FET with the outer/inner electrode as the Source/Drain (left), and the inner/outer electrode as the Source/Drain (right): (b) *n*-region; (e) *p*-region. The differential conductance of the graphene FET with the outer/inner electrode as the Source/Drain (left), and the inner/outer electrode as the Source/Drain (right): (c) *n*-region ($V_G$-$V_{Dirac}$ = 52 V); (f) *p*-region ($V_G$-$V_{Dirac}$ = -68 V).

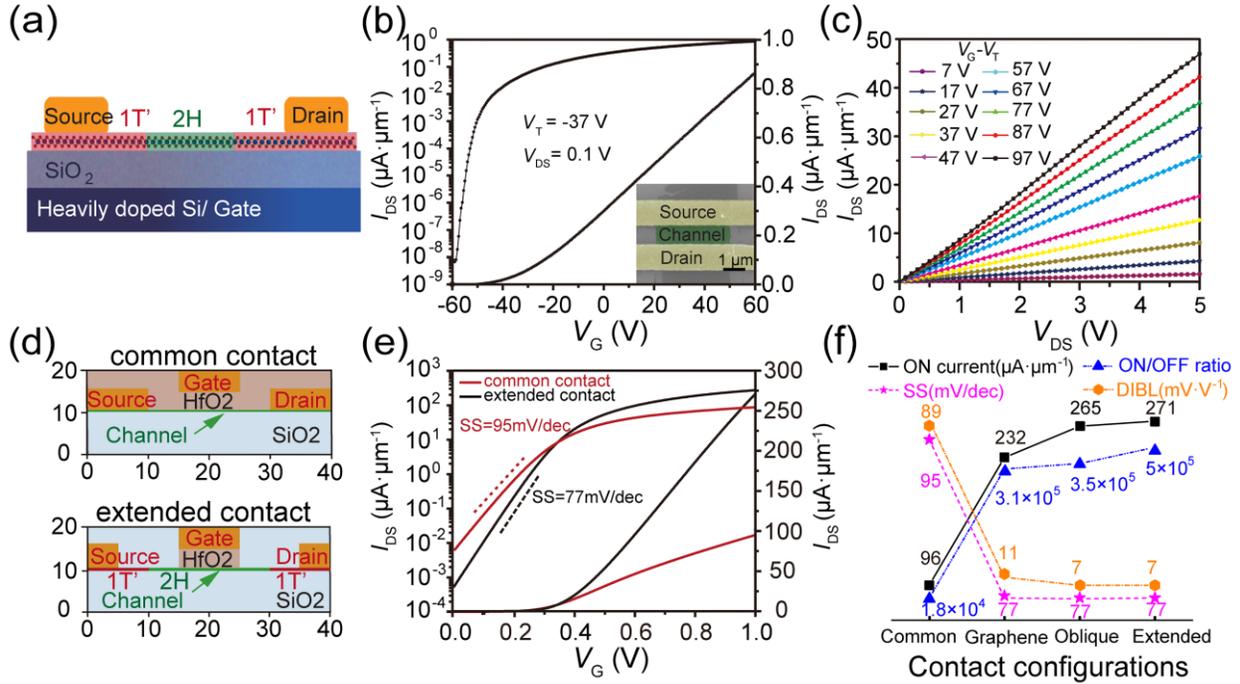

**FIG 4.** Electrical Contact Configuration Engineering. (a) Schematic of the $MoS_2$ FET with the extended Source/Drain electrode. (b) Transfer curve and (c) output curves of the FET in (a). The inset in (b) is the SEM image of the device, the channel length is 0.9 μm. (d) Schematic of the short channel monolayer $MoS_2$ FET with the common contact configuration (upper) and with the optimized contact configuration with extended electrodes/ low dielectric spacer (lower). The contact resistances are both set 800 Ω·μm. (e) Transfer curves of the FETs with the common contact configuration (red) and the extended contact configuration (black). (f) ON current, ON/OFF ratio, SS and DIBL of four short channel $MoS_2$ FETs, with the common contact configuration, the graphene contact configuration, the oblique angle contact configuration, and the extended contact configuration, respectively.

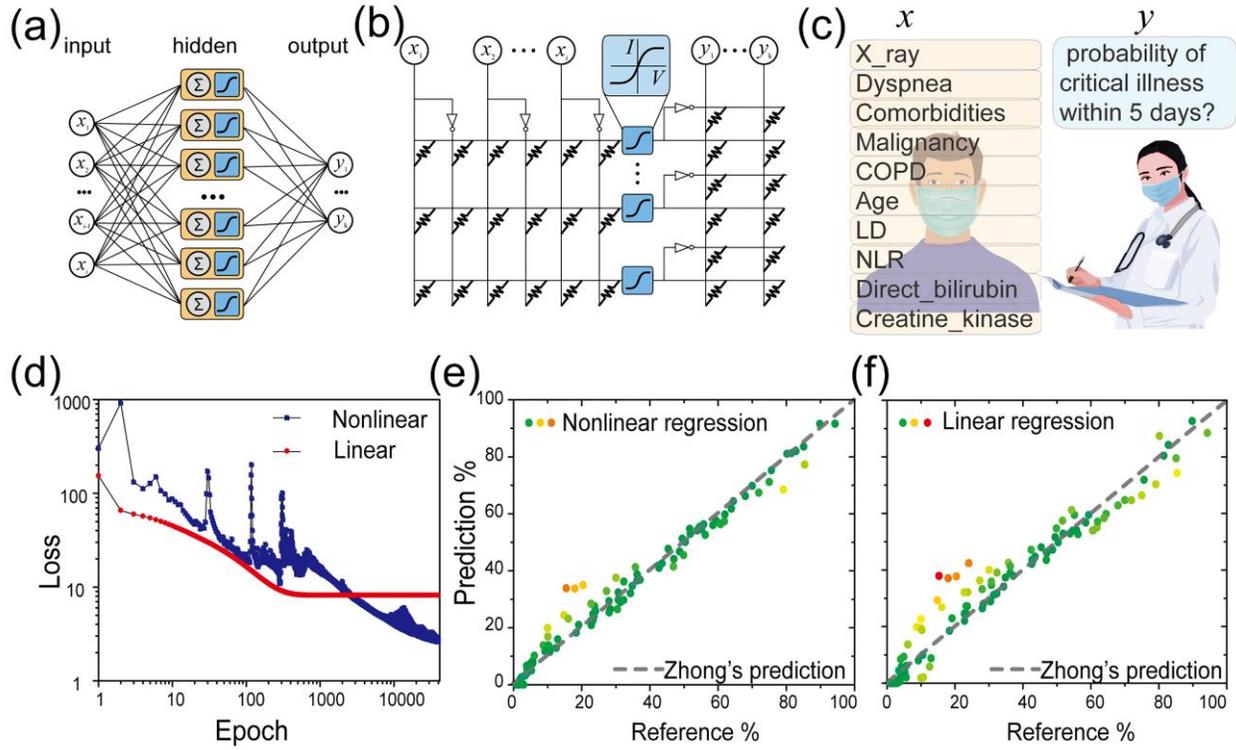

**FIG 5.** Enhancing perceptron ability in ANN circuits with nonlinearity. (a) Schematic of a three-layer ANN. (b) Circuit diagram of the ANN in-memory-computing. (c) Schematic of the COVID-19 early-stage prediction model. COPD: Chronic obstructive pulmonary disease. LD: Lactate dehydrogenase. NLR: Neutrophil/lymphocytes ratio. (d) Loss in the training process. The predicted probability (e) with and (f) without nonlinearity in the ANN circuits. The dotted reference lines are from Zhong et al.'s model described in their earlier paper.

**Table 1**. Devices' performance with various contact configuration

| Contact configuration | Common | | Graphene | | Oblique | | Extended | |
|---|---|---|---|---|---|---|---|---|
| | value | ratio | value | ratio | value | ratio | value | ratio |
| ON current ($\mu A \cdot \mu m^{-1}$) | 96 | - | 232 | 2.4 | 265 | 2.8 | 271 | 2.8 |
| SS (mV/dec) | 94.6 | - | 77 | 0.81 | 76.9 | 0.81 | 77 | 0.81 |
| On/OFF ratio | $1.8 \times 10^4$ | - | $3.1 \times 10^5$ | 17 | $3.5 \times 10^5$ | 19 | $5 \times 10^5$ | 28 |
| DIBL ($mV \cdot V^{-1}$) | 89 | - | 11 | 0.12 | 7 | 0.08 | 7 | 0.08 |

# Supplementary Material

## Field-Effect at Electrical Contacts to Two-Dimensional Materials

Yan Sun[1], Alvin Tang[2], Ching-Hua Wang[2], Yanqing Zhao[1], Mengmeng Bai[1], Shuting Xu[1], Zheqi Xu[1], Tao Tang[3], Sheng Wang[4], Chenguang Qiu[4], Kang Xu[5], Xubiao Peng[1], Junfeng Han[1], Eric Pop[2], and Yang Chai[5], Yao Guo[1*]

[1]School of Physics, Beijing Institute of Technology, Beijing 100081, China.

[2]Department of Electrical Engineering and Stanford SystemX Alliance, Stanford University,
Stanford, California 94305, United States.

[3]Advanced Manufacturing EDA Co., Ltd, Shanghai, 201204, China.

[4]Key Laboratory for the Physics and Chemistry of Nanodevices, Department of Electronics,
Peking University, Beijing 100871, China.

[5]Department of Applied Physics, The Hong Kong Polytechnic University, Hong Kong, China.

*Corresponding author: yaoguo@bit.edu.cn

# I. TCAD simulation of the MoS₂ device with/without the capacitive MIS field-effect

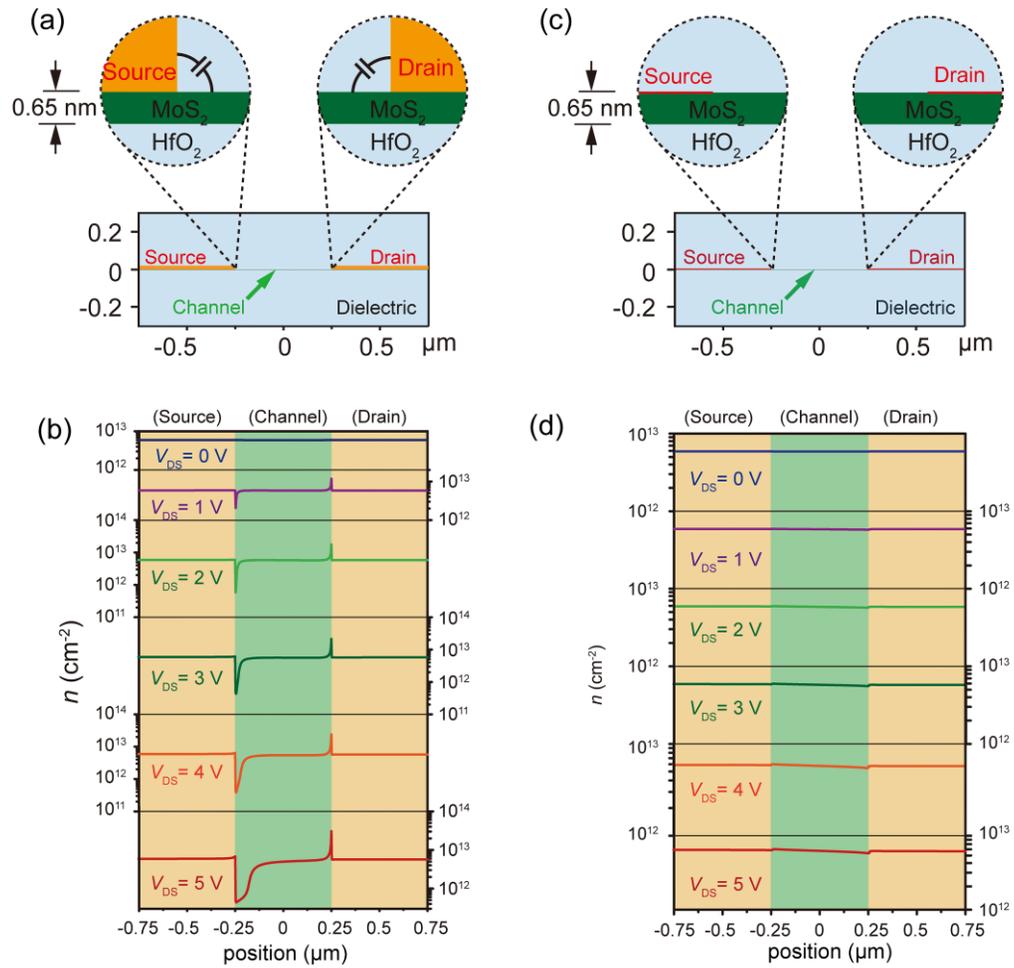

**Figure S1. MoS₂ Devices and carrier densities.** Schematic of the MoS₂ device with (a) and without (c) the sidewall of the metallic electrode. Electron density distribution of the MoS₂ channel with (b) and without (d) the capacitive MIS field-effect.

## II. Capacitive MIS field-effect to Schottky electrical contacts

The manuscript has dicussed the principle of the capacitive MIS field-effect at Ohmic electrical contacts. A similar analysis applies to the Schottky contacts, too. In this case, we carry the same simulation and change the Ohmic contacts to the Schottky contacts with the barrier height of 300 meV. Figs. 2(a) shows the *I-V* curves without the capacitive MIS field-effect. The *I-V* curves are nonlinear with increased differential conductance. This nonlinearity is due to the increased tunneling current, which includes the direct field emission and the thermal-field emission. The large $V_{DS}$ decrease the barrier width and increases the tunneling probability significantly. For the Schottky contacts with the capacitive MIS field-effect, as shown in Figs. 2(b), the *I-V* curves are nonlinear with current saturation. The capacitive MIS field-effect at the Source raises the conduction band energy, increases the barrier width, and therefore reduces the tunneling probability. The band graphs of the two cases are shown in Figs. 2(c) (without the capacitive MIS field-effect) and Figs. 2(d) (with the capacitive MIS field-effect), and zoomed in Figs. 2(e). The barrier width for the direct tunneling increase from 0.2 Å to 2.7 Å due to the capacitive MIS field-effect. The carrier densities are shown in Figs. 2(f).

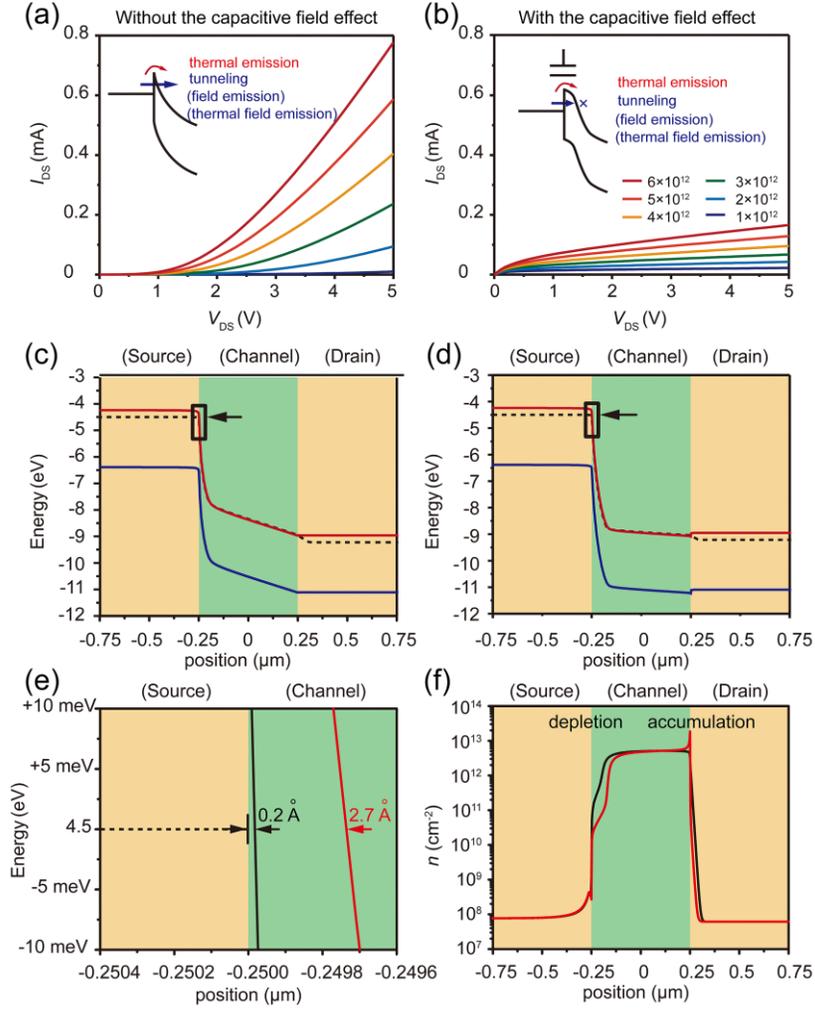

**Figure S2. Capacitive MIS field-effect to Schottky electrical contacts.** The $I_{DS}$-$V_{DS}$ curves of the simulated device using the Schottky barrier of 0.3 eV without (a) and with (b) the capacitive MIS field-effect, the doping density of channel are shown in the legend of (b). Band graph of the device without (c) and with (d) the capacitive MIS field-effect, the solid red lines are the conduction bands, the solid blue lines are the valence bands, and the black dotted lines are the Fermi levels. $V_{DS}$=5 V, the doping density of the channel is $6\times10^{12}$ cm$^{-2}$. (e) Zoomed band graph at the Source end, the black/red lines are the conduction band without/with the capacitive MIS field-effect. The barrier width for the direct tunneling increase from 0.2 Å to 2.7 Å due to the capacitive MIS field-effect. (f) Electron density of the channel, the black/red lines are without/with the capacitive MIS field-effect, respectively.

## III. MoS$_2$ FET with Ni contacts

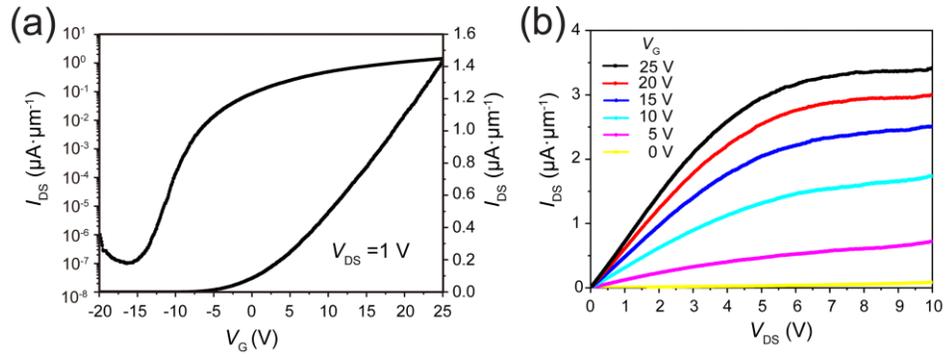

**Figure S3. MoS$_2$ FET with Ni contacts.** Transfer curve (a) and *I-V* curves (b) of the MoS$_2$ FET with Ni contacts. The back gate dielectric is 60 nm thick SiO$_2$, and the channel is CVD grown monolayer MoS$_2$. The channel length is 0.25 μm.

## IV. TEM cross-section images of the electrical contacts to thin channels

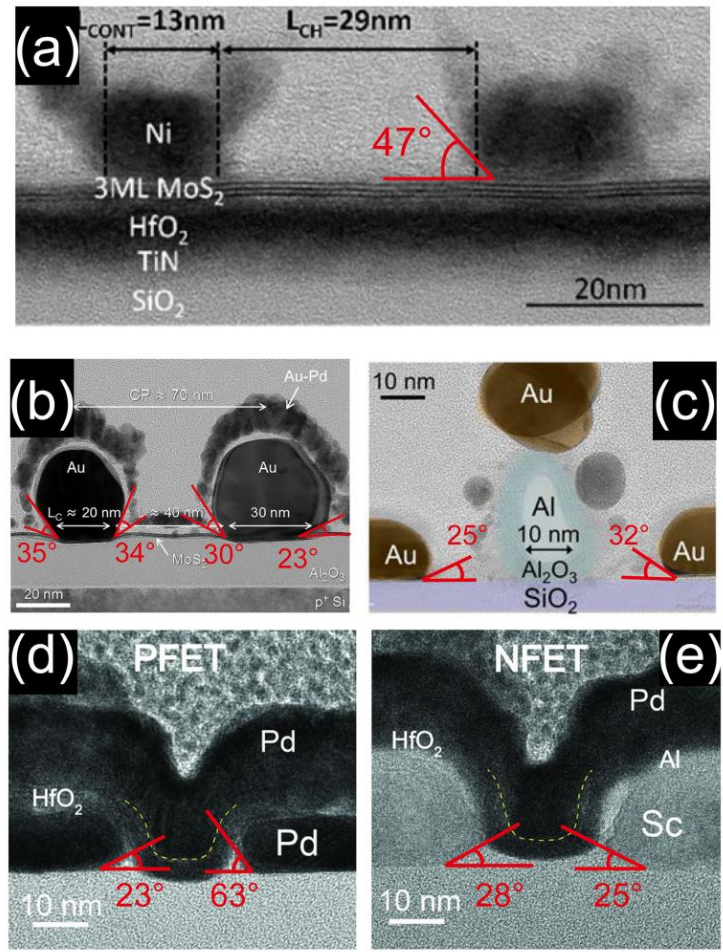

**Figure S4. TEM images of the contacting angle of real devices.** Cross-section of the Ni-MoS$_2$ contact (a),[1] the Au-MoS$_2$ contact (b)[2] and (c),[3] the Pd-carbon nanotube contact (d), and the Sc-carbon nanotube contact (e).[4]

Figure reproduced with permission from: 1, © 2019 IEEE; 2, © 2016 ACS; 3, © 2016 IEEE; 4, © 2017 AAAS;

## V. Enhanced capacitive MIS field-effect with the slanted sidewall

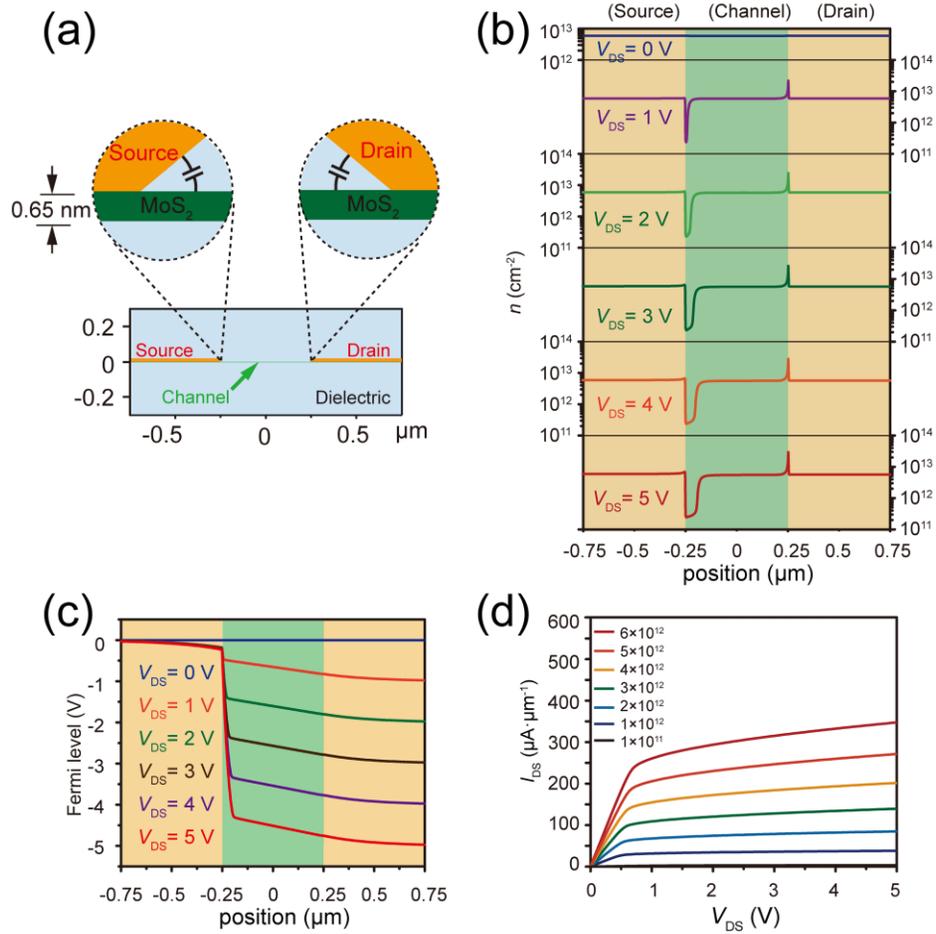

**Figure S5. Enhanced capacitive MIS field-effect with the slanted sidewall.** (a) Schematic of the MoS$_2$ device with the slanted sidewall. (b) Electron density of the MoS$_2$ channel. (c) Voltage potential along the MoS$_2$ channel. (d) Simulated $I_{DS}$-$V_{DS}$ curves, the current saturation is significantly enhanced compared to the results in Fig. 1(f).

## VI. Degeneration of MoS₂ transistors

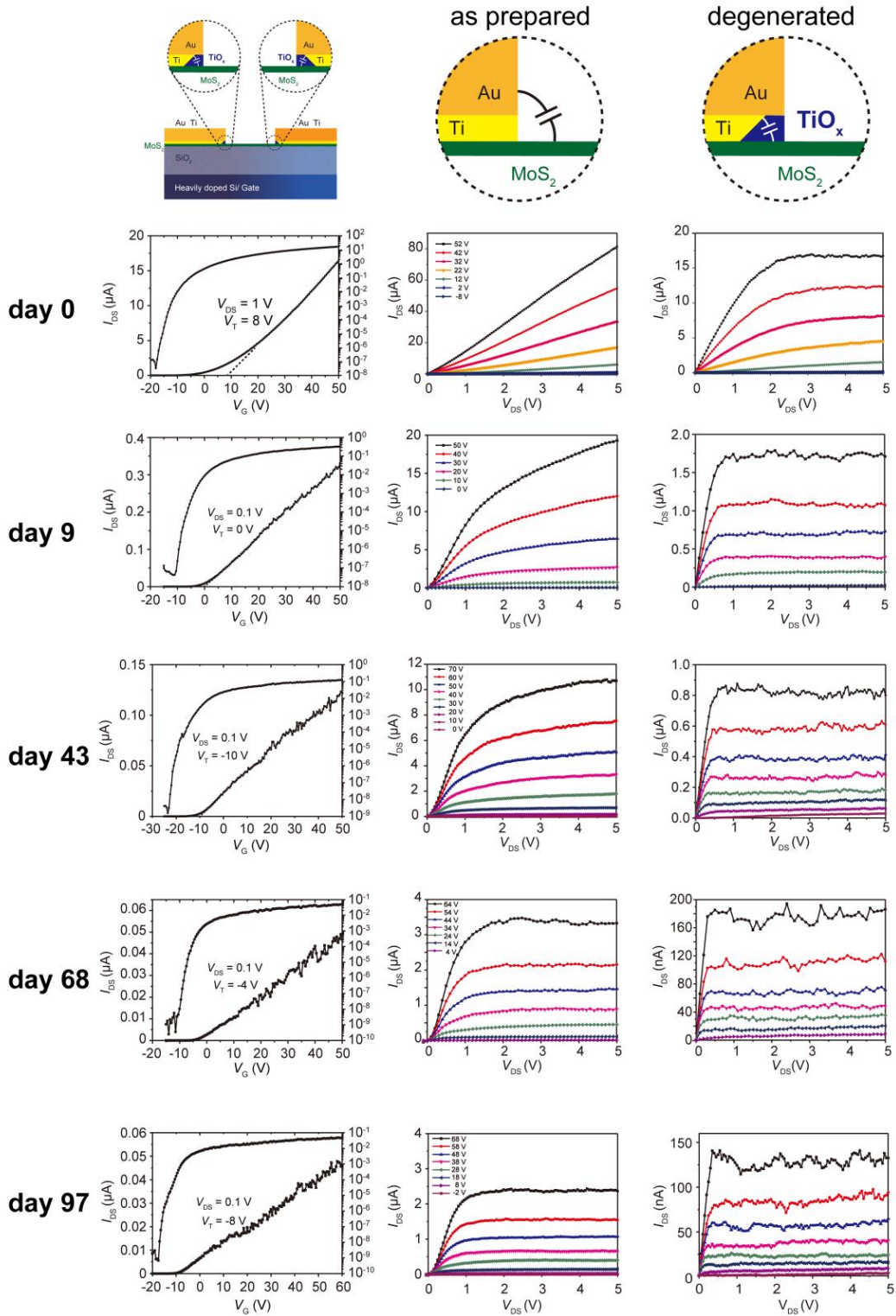

**Figure S6. Degeneration of MoS₂ transistors.** The degeneration of MoS$_2$ transistors is associated with the enhancement of the current saturation. As MoS$_2$ is super stable to the exposure of oxygen, the degeneration should result from the oxygen of Ti to TiO$_x$.

## VII. Capacitive MIS field-effect of the electrical contacts to graphene

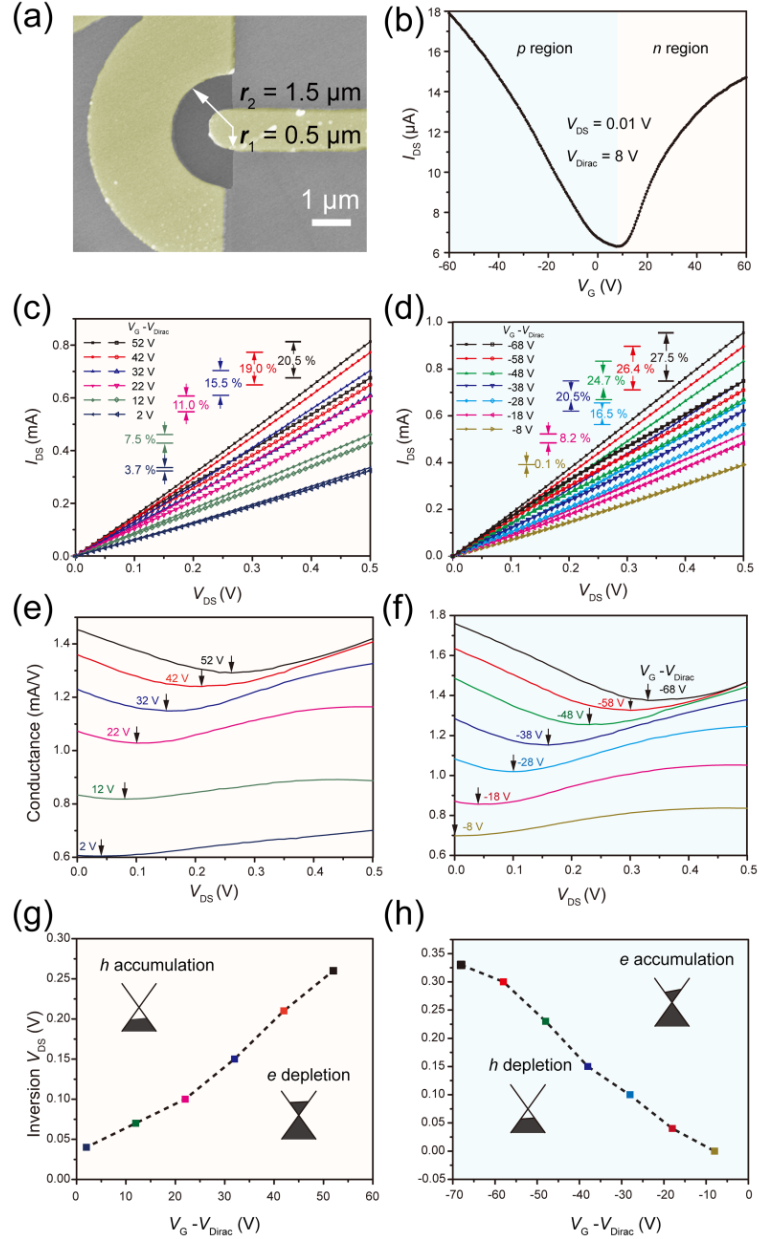

**Figure S7. Capacitive MIS field-effect of the electrical contacts to graphene.** (a) SEM image of the graphene FETs with arc channel. (b) Transfer curve of the graphene FET at $V_{Dirac} = 8$ V. (c) Compared $I_{DS}$-$V_{DS}$ curves with the outer/inner electrode used as the Source/Drain (solid dot) and with the inner/outer electrode used as the Source/Drain (hollow dot). The channel is electron dominated (*n* region, $V_G > V_{Dirac}$). (d) Compared $I_{DS}$-$V_{DS}$ curves with the outer/inner electrode used as the Source/Drain (solid dot) and with the inner/outer electrode used as the Source/Drain (hollow dot). The channel is hole dominated (*p* region, $V_G < V_{Dirac}$). (e) Differential conductance (smoothed) obtained from the *n* region $I_{DS}$-$V_{DS}$ curve in Fig. 3(b) right, with the inner/outer electrode used as the Source/Drain. (f) Differential conductance (smoothed) obtained from the *p* region $I_{DS}$-$V_{DS}$ curve in Fig. 3(e) right, with the inner/outer electrode used as the Source/Drain. (g-h) The inversion points extracted from (e) and (f), respectively.

## VIII. The fabrication process of the contacts with eliminated capacitive MIS field-effect

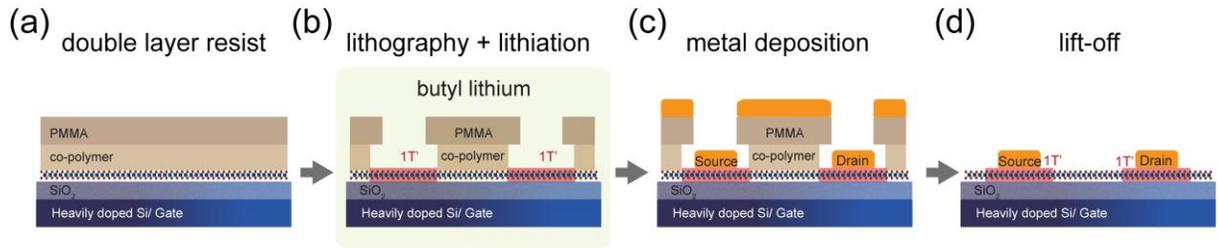

**Figure S8. Schematic of device fabrication.** (a) Double-layer resists of PMMA (up), and co-polymer (down) were used for EBL, which forms the extended space under the pattern. (b) The wafer was then immersed in butyl lithium for the 2H to 1 T' transition of $MoS_2$. (c) The contacting electrodes are then deposited by the e-beam evaporation. (d) The resist and the redundant metal are removed by the lift-off process in acetone.

## IX. 10 nm gate length MoS₂ transistors with different contact configurations

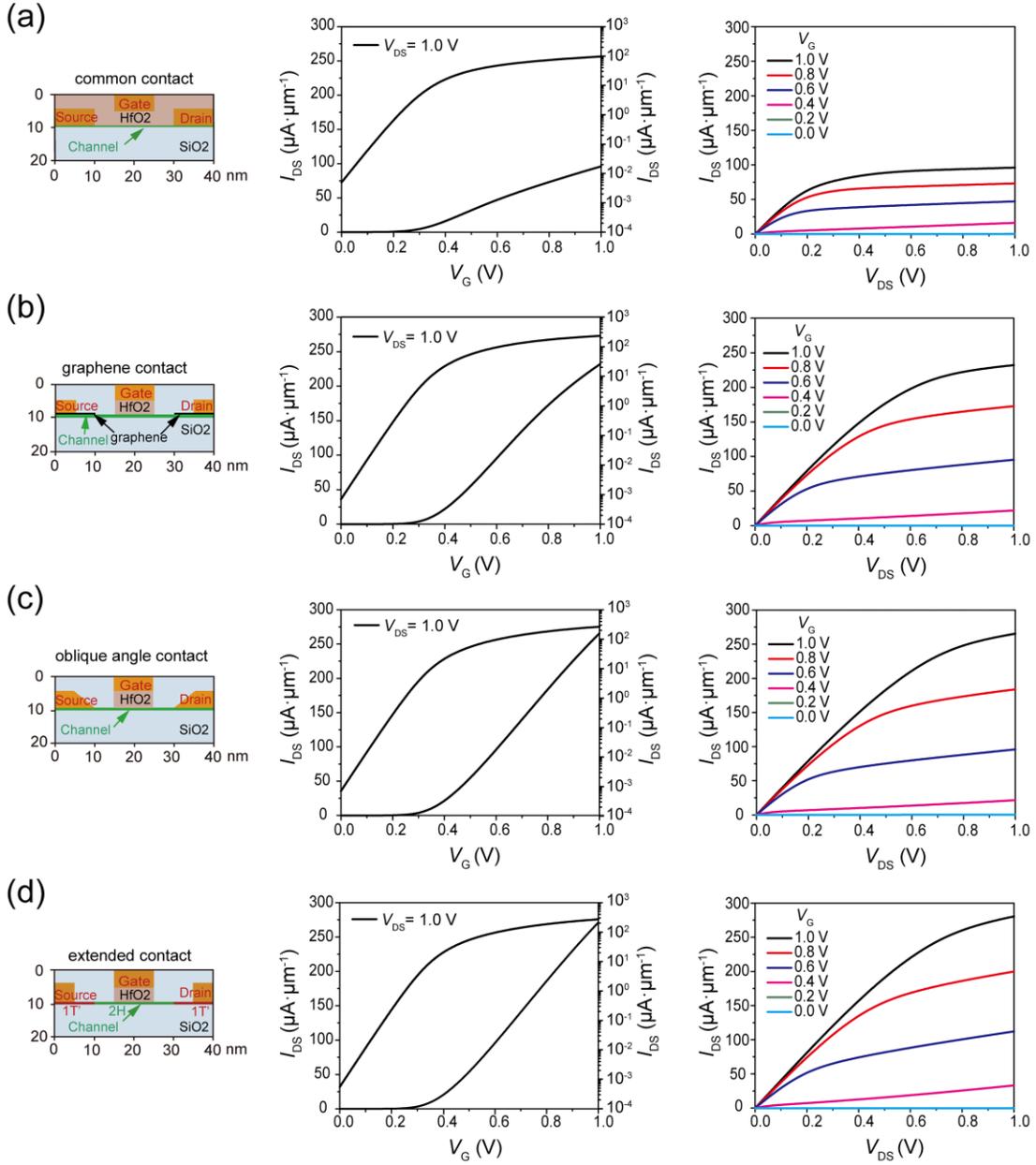

**Figure S9. 10 nm gate length MoS₂ transistors and their electrical characteristics.** Schematic of the short channel monolayer MoS$_2$ FETs with the common contact configuration (a), the graphene contact configuration (b), the oblique angle contact configuration (c), and the extended contact configuration (d). The doping density of the channel is $6\times10^{19}$ cm$^{-2}$. The threshold voltage is adjusted by the work function of the gate electrode (4.9 eV).

## X. Introducing nonlinearity to ANN circuits

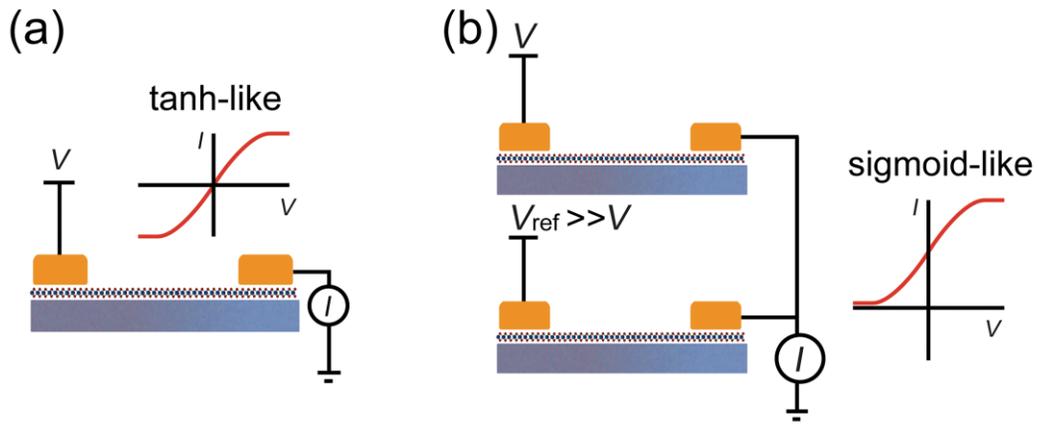

**Figure S10. S-like IV curves by MoS$_2$ devices.** (a) Tanh-like and (b) sigmoid-like IV by simple circuits.

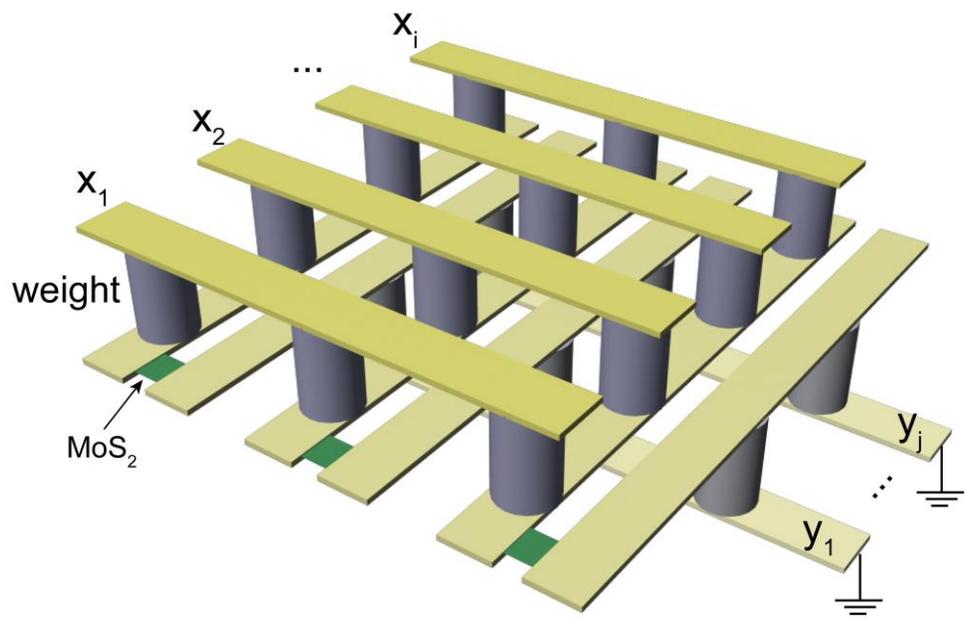

**Figure S11. Compact structure of the hybrid ANN in-memory-computing circuit structure.**